\documentclass[runningheads]{llncs}

\usepackage{pifont}      % 用于 \cmark 和 \xmark 的特殊符号
\usepackage{multirow}    % 用于表格中的多行合并
\usepackage{booktabs}    % 用于 \toprule, \midrule 等漂亮的表格线
\usepackage{algorithm}   % 用于算法伪代码环境
\usepackage{algorithmic} % 用于算法逻辑
\usepackage{marvosym}

\usepackage[T1]{fontenc}
\usepackage{graphicx}
\usepackage{amsmath}
\usepackage{amssymb}
\usepackage{listings}
\usepackage{xcolor}
\usepackage{tikz}
\usetikzlibrary{shapes,arrows,positioning,fit,calc,backgrounds,shadows}
\usepackage{standalone}

\begin{document}

\title{NeuroSCA: Neuro-Symbolic Constraint Abstraction for Smart Contract Hybrid Fuzzing}
\titlerunning{NeuroSCA}

\author{Haochen Liang\inst{1} \and
Jiawei Chen\inst{1} \and
Hideya Ochiai\inst{1}\textsuperscript{(\Letter)}}
\authorrunning{H. Liang et al.}
\institute{Graduate School of Information Science and Technology, The University of Tokyo, Tokyo, Japan\\
\email{\{lianghc,jwchen,ochiai\}@g.ecc.u-tokyo.ac.jp}}

\maketitle

\begin{abstract}
Hybrid fuzzing combines greybox fuzzing’s throughput with the precision of symbolic execution (SE) to uncover deep smart contract vulnerabilities. However, its effectiveness is often limited by constraint pollution: in real-world contracts, path conditions pick up semantic noise from global state and defensive checks that are syntactically intertwined with, but semantically peripheral to, the target branch, causing SMT timeouts. We propose \emph{NeuroSCA} (\emph{Neuro-\textsc{S}ymbolic \textsc{C}onstraint \textsc{A}bstraction}), a lightweight framework that selectively inserts a Large Language Model (LLM) as a semantic constraint abstraction layer. NeuroSCA uses the LLM to identify a small core of goal-relevant constraints, solves only this abstraction with an SMT solver, and validates models via concrete execution in a verifier-in-the-loop refinement mechanism that reintroduces any missed constraints and preserves soundness. Experiments on real-world contracts show that NeuroSCA speeds up solving on polluted paths, increases coverage and bug-finding rates on representative hard contracts, and, through its selective invocation policy, achieves these gains with only modest overhead and no loss of effectiveness on easy contracts.

\end{abstract}

\vspace{-1em}
% ==========================================
\section{Introduction}
\label{sec:intro}
% ==========================================

Smart contracts~\cite{vidal2024vulnerability} on blockchains such as Ethereum manage billions of dollars in digital assets~\cite{liang2024kdtss}. Their immutability makes automated vulnerability discovery a critical priority. Among existing techniques, hybrid fuzzing~\cite{jiang2023evaluating} has become a widely used approach for deep vulnerability detection: it combines the high throughput of greybox fuzzing~\cite{pham2019smart} with the precision of symbolic execution (SE)~\cite{baldoni2018survey}, using fuzzing to explore shallow paths while SE solves hard path conditions that block deeper exploration~\cite{gai2025llama}. This combination allows tools to bypass checks (e.g., magic constants) that purely random fuzzing would likely miss.

However, as decentralized finance (DeFi)~\cite{alamsyah2024review} protocols grow more complex, hybrid fuzzers increasingly suffer from constraint pollution~\cite{wang2024efficiently}. When the symbolic engine targets a specific branch, such as a rare bug oracle deeply nested within protocol logic, it accumulates a path condition $\phi$ as the conjunction of all constraints along the execution trace. In real-world contracts, $\phi$ is rarely clean: it is typically polluted with global invariants that share storage locations or control-flow with the target variables but are only weakly related to the immediate goal of flipping a particular branch.

For instance, consider a critical execution path guarded by a conjunction of heterogeneous constraints, such as complex arithmetic puzzles and environmental invariants like timestamp checks. Ideally, these distinct constraint types would be decoupled to prioritize the resolution of the core logic; however, standard SMT solvers~\cite{winterer2020validating} process a monolithic, coupled formula. Consequently, the solver must address the joint satisfiability of the entire conjunction, often conflating non-linear arithmetic with array theory. This unnecessary coupling significantly expands the search space, leading to frequent timeouts that impede the detection of deep vulnerabilities guarded by such paths~\cite{mikek2023speeding}.

To address this, we propose \emph{NeuroSCA}, a framework that introduces a semantic abstraction layer into the hybrid fuzzing loop. Our approach is grounded in the observation that the semantic intuition of Large Language Models (LLMs) can effectively compensate for the limitations of syntactic solvers~\cite{chen2025cryptic}. While LLMs excel at distinguishing core business logic from defensive noise~\cite{liao2025augmenting}, they lack arithmetic precision; conversely, SMT solvers offer mathematical rigour but lack semantic discrimination~\cite{sun2023smt}. NeuroSCA exploits this complementarity by employing an LLM to identify a small ``semantic core'' of the path condition, which effectively functions as a form of semantic slicing~\cite{wu2023learning}. The system passes only this abstraction, combined with a minimal set of anchored constraints, to the SMT solver. To ensure correctness, NeuroSCA wraps this abstraction in a verifier-in-the-loop mechanism where every model obtained is validated by concrete execution on the Ethereum Virtual Machine (EVM)~\cite{hildenbrandt2018kevm}. If the execution diverges from the intended path, the system identifies the missing constraint and iteratively refines the abstraction until a sound test input is found or the abstraction converges to the full formula.

\paragraph{\textbf{Related Work.}}
Hybrid fuzzers such as ConFuzzius~\cite{torres2021confuzzius} and Smartian~\cite{choi2021smartian} prioritize paths based on control-flow feedback, but inherit SE's scalability limitations when constraints become bloated. "Static slicing techniques~\cite{gallagher2025program,olmedo2025slicing} utilize data-dependency analysis to reduce formula size, yet they are inherently syntactic. In the context of DeFi contracts with pervasive global state~\cite{deng2024safeguarding}, such methods tend to preserve constraints that are semantically peripheral to the target branch. Recently, several works have utilized LLMs to enhance fuzzing, providing new solutions for smart contract vulnerability detection.

\paragraph{\textbf{Outline.}}
Section~\ref{sec:method} presents the NeuroSCA framework, Section~\ref{sec:eval} evaluates its effectiveness, and Section~\ref{sec:conclusion} concludes.

\vspace{-1em}
% ==========================================
\section{The NeuroSCA Framework}
\label{sec:method}
% ==========================================

\begin{figure}[t]
    \centering
    \includegraphics[width=0.95\linewidth]{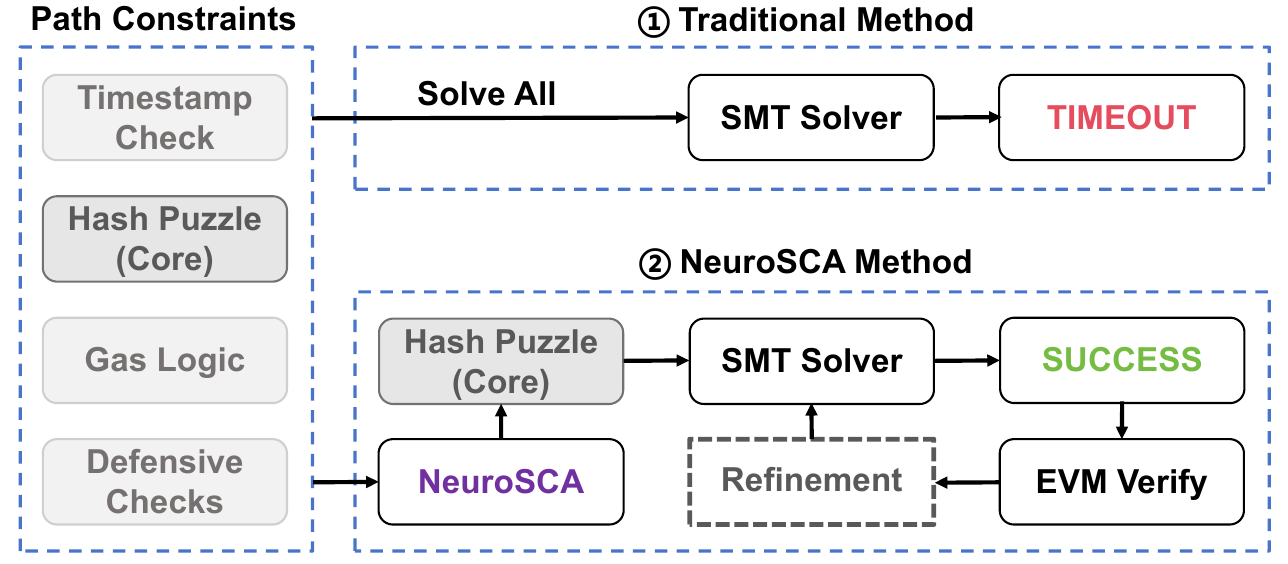}
    \vspace{-1em} 
    \caption{Comparison of constraint solving strategies.}
    \label{fig:arch}
    \vspace{-1.5em}
\end{figure}

As shown in Fig.~\ref{fig:arch}, NeuroSCA is a lightweight backend between symbolic execution (SE) and the SMT solver in a hybrid smart-contract fuzzer. It uses an LLM to build compact, branch-centered abstractions for hard paths and validates solutions via concrete EVM execution in an abstraction–refinement loop, while easy paths are solved by the baseline solver.

\subsection{Architecture and Adaptive Invocation}

Symbolic execution of a seed input $x$ along a path $\pi$ yields a path condition
\begin{equation}
  \phi_\pi = \bigwedge_{i=1}^{n} c_i,
  \label{eq:path-cond}
\end{equation}
where $c_n$ is the guard of the current \texttt{JUMPI} and $c_1,\dots,c_{n-1}$ form its prefix. For a branch-flip goal $g$ the corresponding formula is
\begin{equation}
  \phi_\pi^g = (c_1 \land \dots \land c_{n-1}) \land \neg c_n.
  \label{eq:goal-formula}
\end{equation}

NeuroSCA is integrated as an alternative \texttt{solver-mode} in a ConFuzzius-style fuzzer. Whenever SE generates $\phi_\pi^g$, the system first probes it with a standard SMT solver under a short timeout $T_{\mathit{short}}$. If the probe returns \textsc{sat} or \textsc{unsat}, the baseline result is used; only when it times out or reports \texttt{unknown} is the path classified as hard and handed to NeuroSCA.

To avoid repeated decisions on similar paths, NeuroSCA maintains a cache keyed by a path fingerprint $\mathit{fp}(\phi_\pi^g, g)$ that summarizes the constraint sequence and goal location. Each fingerprint stores a small difficulty label and, when available, a reusable abstraction, so repeated fingerprints can directly select the solver mode without new LLM queries.

\vspace{-1em}
\subsection{Anchored Constraint Abstraction}

Given a hard path condition and goal, NeuroSCA first performs an anchored abstraction that separates mandatory constraints from those that can be subjected to neuro-symbolic selection and refinement. For a branch-flip goal we write
\begin{equation}
  \phi_\pi^g = \phi_{\mathit{pref}} \land \neg c_n,
  \label{eq:goal-split}
\end{equation}
where $\phi_{\mathit{pref}} = c_1 \land \dots \land c_{n-1}$. NeuroSCA decomposes
\begin{equation}
  \phi_{\mathit{pref}} = \phi_{\mathit{hard}} \land \phi_{\mathit{soft}},
  \label{eq:pref-split}
\end{equation}
where $\phi_{\mathit{hard}}$ contains the essential backbone of the path and $\phi_{\mathit{soft}}$ collects the remaining constraints.

Concretely, $\phi_{\mathit{hard}}$ keeps (i) definitions and data dependencies of the guard $c_n$; (ii) the function selector and basic input-shape constraints such as calldata layout; and (iii) minimal safety checks needed to avoid malformed transactions or immediate crashes. The rest is placed into $\phi_{\mathit{soft}}$, which often contains global-state checks and defensive assertions that dominate solving time but are only weakly related to the current goal. The partition is implemented via simple local control-flow and data-flow rules; additional examples are given in the appendix.

\vspace{-1em}
\subsection{LLM-Guided Core Selection and Heuristic Fallback}

Given $\phi_{\mathit{soft}}$, NeuroSCA selects a small core of constraints that, together with $\phi_{\mathit{hard}}$ and the negated guard, still suffices to guide the solver. Each constraint $c_i \in \phi_{\mathit{soft}}$ is serialized into a readable string with lightweight metadata. NeuroSCA builds a compact prompt summarizing the goal $g$ and the local path context and presents the annotated constraints as a numbered list. The LLM returns indices of a small set of semantically most critical constraints together with an ordered ranking of all candidates. Let $K \subseteq \{ i \mid c_i \in \phi_{\mathit{soft}} \}$ denote the selected core; the initial abstracted formula is
\begin{equation}
  \psi_0 = \phi_{\mathit{hard}} \land \bigwedge_{i \in K} c_i \land \neg c_n.
  \label{eq:psi0}
\end{equation}
In our implementation, $|K|$ is bounded by a small constant and the returned indices are checked for range and format. When the external LLM is unavailable or a call fails, NeuroSCA falls back to a structural heuristic that selects the $K$ most recent constraints from $\phi_{\mathit{soft}}$, based on the intuition that constraints closest to the guard most directly influence it. Successful core sets are recorded in the fingerprint cache and reused whenever the same pattern reappears.
\vspace{-1em}
\subsection{Verifier-in-the-Loop Abstraction--Refinement}

LLM-guided abstraction may over-approximate the path condition, omitting constraints that are logically necessary so that models of $\psi_0$ fail on the real program. NeuroSCA mitigates this via a verifier-in-the-loop procedure that alternates SMT solving with concrete EVM execution.

Given the current abstraction $\psi_k$ (initially $\psi_0$), NeuroSCA calls the SMT solver on $\psi_k$ with timeout $T_{\mathit{neur}}$. If a model $M_k$ is found, it is concretized into an input $x_k$ and executed on an instrumented EVM, yielding a branch trace $\tau_k$ that is compared with the symbolic path of $\phi_\pi^g$. If execution follows that path and reaches $g$, $x_k$ is accepted as a new seed; otherwise NeuroSCA locates the first divergent branch, retrieves the corresponding missing constraint $c_{\mathit{miss}}$ from the original path condition, and refines the abstraction:
\begin{equation}
  \psi_{k+1} = \psi_k \land c_{\mathit{miss}}.
  \label{eq:refine}
\end{equation}

The loop repeats up to a refinement bound $R$, after which NeuroSCA abandons the path or optionally falls back to solving the full $\phi_\pi^g$ with a longer timeout. In the worst case, refinement reconstructs the full formula and the backend degenerates to baseline solving; on typical polluted paths only a few constraints are reintroduced before a feasible model is obtained.

\vspace{-1em}
% ==========================================
\section{Evaluation}
\label{sec:eval}
% ==========================================

We evaluate NeuroSCA from two perspectives:
\emph{RQ1}---micro-level efficiency on polluted path conditions; and
\emph{RQ2}---bug-finding capability on challenging contracts.
We compare three configurations: the baseline solver (Baseline), always applying NeuroSCA (NeuroSCA-only), and our selective method (Selective).

\vspace{-0.8em}
\subsection{RQ1: Micro-level Efficiency on Polluted Paths}

For RQ1 we analyse per-call solver statistics in Table~\ref{tab:combined}.
\texttt{DeepConstraintTime\-lock} and \texttt{MultiStepStatefulPuzzle} are synthetic stress tests with heavily polluted guards; the remaining contracts are real-world or benchmark tokens with mostly lightweight constraints.

On \texttt{DeepConstraintTime\-lock}, NeuroSCA-only sharply reduces solving latency on hard paths: the average per-call time drops from 17.84\,s to 6.25\,s and the P99 tail from 44.80\,s to 22.49\,s.
Selective NeuroSCA still lowers the P99 to 29.11\,s while issuing fewer calls than NeuroSCA-only (178 vs.\ 379), because it switches away from the baseline only when the probe classifies a path as hard.

On the easier contracts (e.g. \texttt{Infinity}, \texttt{SmartBillions}, \texttt{WallOfChainToken}, \texttt{ReclaimTokens}), Selective closely tracks the baseline in both average time and tail (e.g., 0.40\,s vs.\ 0.35\,s on \texttt{Infinity}), whereas NeuroSCA-only pays a much higher P99 (up to 3.40\,s) due to unnecessary abstraction.
This shows that the selective policy avoids neuro-symbolic overhead on simple path conditions.

On \texttt{MultiStepStatefulPuzzle}, NeuroSCA-only and Selective attempt many more guarded branches than Baseline (45 and 171 calls vs.\ 2).
Despite the increased exploration, Selective keeps the average latency close to the baseline (4.77\,s vs.\ 4.45\,s) and clearly improves over NeuroSCA-only (7.06\,s), while slightly tightening the P99 tail.
These extra calls correspond to deeper behaviours that matter for bug finding in RQ2.

\begin{table}[t]
\vspace{-1em}
\centering
\caption{End-to-end fuzzing and per-call solver statistics on representative contracts.}
\label{tab:combined}
\scriptsize
\setlength{\tabcolsep}{3pt}
\begin{tabular}{@{}p{0.25\linewidth}l rr rrr@{}}
\toprule
 & & \multicolumn{2}{c}{Fuzzing metrics} & \multicolumn{3}{c}{Solver metrics} \\
\cmidrule(lr){3-4} \cmidrule(lr){5-7}
\textbf{Contract} & \textbf{Config.} &
\textbf{Cov. (\%)} & \textbf{\#Bugs} &
\textbf{\#Calls} & \textbf{Avg (s)} & \textbf{P99 (s)} \\
\midrule
\multirow{3}{*}{\texttt{Infinity}}
  & Baseline   & 59.8 & 3 & 120 & 0.35 & 1.20 \\
  & NeuroSCA-only & 38.9 & 1 & 340 & 0.95 & 3.40 \\
  & Selective     & \textbf{65.9} & \textbf{3} & 180 & 0.40 & 1.50 \\
\midrule
\multirow{3}{*}{\texttt{MultiStepStatefulPuzzle}}
  & Baseline   & 67.9 & 0 & 2   & 4.45 & 8.89 \\
  & NeuroSCA-only & 62.5 & 0 & 45  & 7.06 & 11.58 \\
  & Selective     & \textbf{80.2} & \textbf{2} & 171 & 4.77 & 10.47 \\
\midrule
\multirow{3}{*}{\texttt{SmartBillions}}
  & Baseline   & 52.9 & 19 & 210 & 0.50 & 2.30 \\
  & NeuroSCA-only & 52.0 & 13 & 480 & 1.10 & 4.80 \\
  & Selective     & \textbf{65.7} & \textbf{21} & 320 & 0.55 & 2.00 \\
\midrule
\multirow{3}{*}{\texttt{WallOfChainToken}}
  & Baseline   & 68.6 & 0 & 90  & 0.30 & 1.00 \\
  & NeuroSCA-only & 67.5 & 1 & 260 & 0.85 & 3.00 \\
  & Selective     & \textbf{71.1} & \textbf{1} & 150 & 0.38 & 1.30 \\
\midrule
\multirow{3}{*}{\texttt{ReclaimTokens}}
  & Baseline   & 61.2 & 2 & 140 & 0.40 & 1.60 \\
  & NeuroSCA-only & 54.1 & 1 & 360 & 0.90 & 3.80 \\
  & Selective     & \textbf{67.4} & \textbf{2} & 200 & 0.42 & 1.70 \\
\midrule
\multirow{3}{*}{\texttt{DeepConstraintTime\-lock}}
  & Baseline   & 97.8 & 8 & 173 & 17.84 & 44.80 \\
  & NeuroSCA-only & 95.0 & 6 & 379 &  6.25 & 22.49 \\
  & Selective     & \textbf{98.3} & \textbf{8} & 178 & 13.11 & 29.11 \\
\bottomrule
\end{tabular}
\vspace{-1.5em}
\end{table}

\vspace{-1em}
\subsection{RQ2: Bug Finding on Synthetic and Real Contracts}

For RQ2 we interpret the fuzzing metrics in Table~\ref{tab:combined}.
Across these contracts, Selective NeuroSCA matches or improves Baseline on coverage and bug count in every case, while typically issuing more solver calls on the harder benchmarks (\texttt{MultiStepStatefulPuzzle}, \texttt{DeepConstraintTime\-lock}).
This indicates that NeuroSCA explores more guarded paths without sacrificing solving efficiency.

On the synthetic \texttt{MultiStepStatefulPuzzle}, Baseline reaches 67.9\% coverage but never triggers the deep oracles.
Selective raises coverage to 80.2\%, turns previously unreachable guarded branches into two concrete bugs, and does so with per-call cost comparable to the baseline.
On the real-world contracts \texttt{Infinity}, \texttt{SmartBillions}, \texttt{WallOfChainToken}, and \texttt{ReclaimTokens}, Selective consistently yields higher coverage and equal or more bugs, whereas NeuroSCA-only spends more solver time yet often loses bugs.

Beyond these examples, we ran NeuroSCA on a larger dataset of 100 real-world contracts (not shown in the table) and observed coverage improvements on 85\% of them, with no statistically significant loss of bug-finding ability.
On simple contracts such as \texttt{ArraySumPuzzle}, \texttt{BasicToken}, \texttt{Ownable}, and \texttt{usingOraclize}, all three configurations behave almost identically, indicating that Selective NeuroSCA effectively degenerates to the baseline when constraints are already easy.

Overall, the results show that NeuroSCA delivers the intended trade-off:
on polluted, vulnerability-guarded paths it substantially strengthens a strong hybrid fuzzer and explores more useful solver queries, while on ordinary paths the selective invocation policy avoids the large overheads of naive ``always-on'' neuro-symbolic integration.

\vspace{-1em}
% ==========================================
\section{Conclusion}
\label{sec:conclusion}
% ==========================================
In this paper, we presented \textsc{NeuroSCA}, a neuro-symbolic framework designed to mitigate constraint pollution in smart contract hybrid fuzzing. By synergizing anchored constraint abstraction with a verifier-in-the-loop refinement mechanism, our approach accelerates the resolution of complex, noise-laden paths while strictly preserving soundness. Empirical results demonstrate that \textsc{NeuroSCA} successfully unlocks deep vulnerability-guarding branches inaccessible to state-of-the-art fuzzers, maintaining high efficiency through its adaptive invocation policy. Future work will focus on extending this neuro-symbolic paradigm to automate invariant inference and broader cross-contract vulnerability detection.

% ==========================================
% REFERENCES
% ==========================================
\bibliographystyle{unsrt}
\bibliography{sample-base}

% ==========================================
% APPENDIX
% ==========================================
\clearpage
\appendix
\section*{Appendix}

% ------------------------------------------
% Section A
% ------------------------------------------
\section{Experimental Setup Details}
\label{app:setup}

We summarise here the configurations used in our evaluation to facilitate reproducibility and to complement Section~\ref{sec:eval}.

\paragraph{Environment.}
All experiments were performed on a server equipped with 16~vCPUs and 32~GB of RAM, running Ubuntu~20.04.
NeuroSCA uses Z3~(v4.8) as the SMT solver and a lightweight GPT-4–class model (GPT-4o-mini) as the Large Language Model backend, accessed via an HTTP API.
Unless otherwise noted, each fuzzing campaign runs for a fixed time budget (30~minutes) with a fixed random seed; all numbers in the main tables are taken from a single representative run, while preliminary experiments on three independent runs showed very small variance.

\paragraph{Solver configurations.}
We compare three strategies:
(1) \emph{Baseline-SE}, the standard ConFuzzius symbolic executor with a 60\,s timeout per path condition;
(2) \emph{NeuroSCA-only}, which applies anchored abstraction and verifier-in-the-loop refinement to every path, using a shorter timeout $T_{\mathit{neur}}$ (5\,s) on abstracted formulas;
and (3) \emph{Selective NeuroSCA}, our adaptive backend that first probes each path condition with the baseline solver (1--2\,s) and only switches to NeuroSCA if the probe fails.
All evolutionary fuzzing parameters (mutation rates, selection strategy, and corpus management) are kept identical across these modes, so differences arise solely from the path-solving backend.

\paragraph{Benchmark corpus.}
Our dataset contains 100 real contracts collected from public repositories.
The contracts in Table~\ref{tab:combined} are a representative subset used for detailed reporting:
they cover different protection patterns (timelocks, multi-step state updates, lottery-style logic, and guarded withdrawals) and illustrate the typical behaviour of the three configurations on the wider corpus.

% ------------------------------------------
% Section B
% ------------------------------------------
\section{Verifier-in-the-Loop Refinement}
\label{app:algorithm}

Algorithm~\ref{alg:refinement_appendix} formalises the core logic of NeuroSCA.
The process begins with an anchored decomposition, separating the goal-augmented path condition $\phi_\pi^g$ into a mandatory hard core $\phi_{\mathit{hard}}$ and a soft candidate set $\phi_{\mathit{soft}}$.
The LLM (or a structural fallback) selects a small core set $K$ from $\phi_{\mathit{soft}}$, yielding the initial abstraction $\psi_0$.

The system then enters an iterative refinement loop.
In each step it solves $\psi_k$ to obtain a model $M_k$, concretises $M_k$ into an input $x_k$, and executes $P$ on an instrumented EVM to obtain a concrete trace $\tau_k$.
If $\tau_k$ follows the intended symbolic path and reaches the goal, $x_k$ is accepted as a sound test input.
Otherwise NeuroSCA finds the first missing constraint $c_{\mathit{miss}}$ on the diverging branch and strengthens the abstraction ($\psi_{k+1} \gets \psi_k \land c_{\mathit{miss}}$).
This ensures that every reported bug is confirmed by concrete execution; in the worst case, the abstraction converges back to the full formula.
In all experiments we bound the number of refinement rounds $R$ to a small constant (typically 3--5), which was sufficient for every successful path.

% --- Algorithm 1 ---
\begin{algorithm}[t!]
\small
\caption{Verifier-in-the-loop refinement in NeuroSCA}
\label{alg:refinement_appendix}
\begin{algorithmic}[1]
\REQUIRE Path condition $\phi_\pi^g$, goal $g$, refinement bound $R$, timeout $T_{\mathit{neur}}$
\STATE $(\phi_{\mathit{hard}}, \phi_{\mathit{soft}}, c_n) \gets \mathsf{AnchoredDecompose}(\phi_\pi^g, g)$
\STATE $(K, \mathit{order}) \gets \mathsf{CoreSelect}(\phi_{\mathit{soft}}, g)$ \COMMENT{LLM or heuristic}
\STATE $\psi_0 \gets \phi_{\mathit{hard}} \land \bigwedge_{i \in K} c_i \land \neg c_n$
\FOR{$k = 0$ \TO $R$}
  \STATE $(\mathit{res}, M_k) \gets \mathsf{SolveSMT}(\psi_k, T_{\mathit{neur}})$
  \IF{$\mathit{res} \neq \text{SAT}$}
    \RETURN \texttt{FAIL} \COMMENT{No model within current abstraction}
  \ENDIF
  \STATE $x_k \gets \mathsf{Concretize}(M_k)$
  \STATE $\tau_k \gets \mathsf{ExecuteInstrumented}(P, x_k)$
  \IF{$\mathsf{ReachesGoal}(\tau_k, g)$ \AND $\mathsf{ConsistentWithPath}(\tau_k, \phi_\pi^g)$}
    \RETURN \texttt{SAT}, $x_k$ \COMMENT{Sound test input}
  \ENDIF
  \STATE $c_{\mathit{miss}} \gets \mathsf{FirstMissingConstraint}(\tau_k, \phi_\pi^g)$
  \IF{$c_{\mathit{miss}} = \bot$}
    \RETURN \texttt{FAIL}
  \ENDIF
  \STATE $\psi_{k+1} \gets \psi_k \land c_{\mathit{miss}}$ \COMMENT{Refine abstraction}
\ENDFOR
\RETURN \texttt{FAIL}
\end{algorithmic}
\end{algorithm}

% --- Table: Anchoring policy ---
\begin{table}[t!]
\centering
\caption{Constraint anchoring policy used in NeuroSCA.}
\label{tab:anchors}
\scriptsize
\setlength{\tabcolsep}{4pt}
\begin{tabular}{@{}p{0.22\linewidth} p{0.55\linewidth} p{0.14\linewidth}@{}}
\toprule
\textbf{Category} & \textbf{Typical form / example} & \textbf{Policy} \\ \midrule
Target guard & Branch predicate at the target location (for example the current \texttt{JUMPI}). & Hard (anchor) \\
Data dependencies & Definitions feeding into the guard (for example arithmetic on arguments, storage loads flowing to $c_n$). & Hard (anchor) \\
Input shape & \texttt{msg.sig}, \texttt{msg.data.length}, ABI layout constraints. & Hard (anchor) \\
Minimal safety & Checks preventing malformed calls or immediate crashes (for example array bounds). & Hard (anchor) \\
\midrule
Timelocks / flags & \texttt{block.timestamp} windows, pause flags, global caps. & Soft (candidate) \\
Unrelated storage & \texttt{SLOAD} from slots not data-dependent on $c_n$. & Soft (candidate) \\
Complex arithmetic & Non-linear arithmetic not flowing to $c_n$ or directly to the goal. & Soft (candidate) \\
\bottomrule
\end{tabular}
\end{table}

\paragraph{Constraint anchoring.}
Table~\ref{tab:anchors} summarises the static rules for partitioning constraints.
By anchoring the target branch guard, its data dependencies, and basic input-integrity checks, we ensure that the LLM only abstracts environmental noise (for example timelocks and global flags) without breaking transaction well-formedness.

% ------------------------------------------
% Section C
% ------------------------------------------
\section{Supplementary Experimental Results}
\label{app:experiments}

\paragraph{Case study}
Table~\ref{tab:case_multistep} analyses a separate micro-benchmark that directly targets the hardest guarded branch in \texttt{MultiStepStatefulPuzzle}, which is protected by global flags and multi-step state changes.
On this branch the Baseline-SE configuration accumulates 91 constraints and typically hits the 60\,s solving timeout without producing a model.
NeuroSCA abstracts the path condition to a core of 34 constraints and, after three refinement steps, solves the branch in roughly 3--4\,s.
Selective NeuroSCA reuses the same abstraction but only on paths classified as hard, achieving essentially identical solving behaviour while still keeping easy paths on the baseline solver.
This case illustrates how anchored abstraction and refinement cut through global-state noise without sacrificing soundness and explains why, in the full fuzzing runs of Table~\ref{tab:combined}, only Selective manages to exercise the deep bug oracles in this contract.

\begin{table}[t!]
\centering
\caption{Micro-level branch-solving behaviour on a guarded path in \texttt{MultiStepStatefulPuzzle}.}
\label{tab:case_multistep}
\scriptsize
\begin{tabular*}{0.95\textwidth}{@{\extracolsep{\fill}} lccccc @{}}
\toprule
\textbf{Config.} & $|\phi_\pi^g|$ & $|\psi_0|$ & \textbf{Solve time} & \textbf{\#Refine} & \textbf{Bug found?} \\
\midrule
Baseline-SE      & 91 & -- & $> 60$\,s (T.O.) & -- & No \\
NeuroSCA-only    & 91 & 34 & 3.4\,s          & 3  & Yes (micro) \\
Selective        & 91 & 34 & 3.2\,s          & 3  & Yes (micro) \\
\bottomrule
\end{tabular*}
\end{table}

\paragraph{Evaluation on Real-World Contracts Dataset.}
We conducted an evaluation on a dataset comprising 100 real-world smart contracts to benchmark \textsc{NeuroSCA} against the \textsc{ConFuzzius} baseline. The results demonstrate that \textsc{NeuroSCA} outperforms the baseline by achieving higher instruction coverage and detecting a greater number of vulnerabilities. These findings confirm the effectiveness of our approach in handling the diverse and complex constraints found in real-world deployments.

\begin{table}[t!]
    \centering
    \caption{Comparison of performance on real-world contracts.}
    \label{tab:simple_comparison}
    \begin{tabular*}{0.95\textwidth}{@{\extracolsep{\fill}} lcc @{}}
        \toprule
        \textbf{Method} & \textbf{Bugs} & \textbf{Instr. Coverage} \\
        \midrule
        ConFuzzius (Baseline) & 106 & 76.4\% \\
        NeuroSCA    & 122 & 82.1\% \\
        \bottomrule
    \end{tabular*}
\end{table}
% ------------------------------------------
% Section D
% ------------------------------------------
\section{LLM Prompt Template}
\label{app:prompts}

Table~\ref{tab:coreselect-prompts} shows the prompt design used in our prototype.
By asking the LLM to rank constraints by semantic relevance, rather than to solve them, NeuroSCA leverages high-level understanding while delegating precise reasoning to the SMT solver.
The JSON output format simplifies validation, caching, and reproducibility.

\begin{table}[t!]
\footnotesize
\centering
\caption{Prompt templates for LLM-guided core selection in NeuroSCA.}
\label{tab:coreselect-prompts}
\renewcommand{\arraystretch}{1.1}
\begin{tabular}{p{0.18\textwidth} p{0.76\textwidth}}
\hline
\textbf{Template} & \textbf{Prompt content} \\
\hline
System role &
You are an expert on smart-contract analysis and symbolic execution.
Given a target branch and a set of path constraints, identify which constraints are semantically most important for deciding whether the branch can be taken.
Do not attempt to solve the constraints; instead, rank them by relevance to the goal. \\[0.4em]

User query &
\textbf{Goal:} We want to satisfy the branch at program counter \texttt{<PC>} in function \texttt{<FUNC>}. The high-level guard of this branch is \texttt{<GUARD\_DESC>}. \newline
\textbf{Constraints:} Below is a list \texttt{<SOFT\_CONSTRAINTS>} of candidate constraints. Each entry has an \texttt{id}, \texttt{expr}, \texttt{kind}, and \texttt{vars}.
Example: \texttt{id=1, expr="block.timestamp > unlockTime", kind="timelock", vars=["block.timestamp","unlockTime"]}. \newline
\textbf{Task:} (1) Choose a small list \texttt{core\_indices} (typically 4--16 ids) that are most important for controlling whether the goal branch can be taken.
(2) Produce a complete \texttt{ranking} of all constraint ids from most to least important. \newline
\textbf{Guidance:} Treat generic timelocks and global flags as noise unless they clearly dominate feasibility.
Focus on constraints that directly affect balances, user inputs, and variables in the guard. \newline
\textbf{Output format:} A JSON object of the form \texttt{\{"core\_indices": [...], "ranking": [...]\}} only. \\
\hline
\end{tabular}
\end{table}

\end{document}